\documentclass{article}
\usepackage{spconf,amsmath,graphicx,algorithm,algorithmic,adjustbox,setspace}

\usepackage{xcolor}
\usepackage{enumitem}
\usepackage{hyperref}
\usepackage{multirow}
\usepackage{etoolbox,siunitx}
\robustify\bfseries
\usepackage{booktabs}
\usepackage{balance}
\usepackage{amssymb}
\usepackage{bm}
\usepackage{dsfont}
\usepackage{cite}

\usepackage[hang,flushmargin]{footmisc}

\DeclareMathOperator*{\sigmoid}{sigmoid}
\DeclareMathOperator*{\gelu}{GeLU}
\usepackage[inkscapelatex=false]{svg}

\let\OLDthebibliography\thebibliography
\renewcommand\thebibliography[1]{
  \OLDthebibliography{#1}
  \setlength{\parskip}{.6pt}
  \setlength{\itemsep}{.6pt plus 0.3ex}
}

\title{NIIRF: Neural IIR Filter Field for HRTF Upsampling and Personalization}
\name{{\shortstack[c]{Yoshiki Masuyama$^{1,2}$, Gordon Wichern$^{1}$, François G.\ Germain$^{1}$, Zexu Pan$^{1}$,\\ Sameer Khurana$^{1}$, Chiori Hori$^1$, Jonathan Le Roux$^{1}$\thanks{This work was performed while Y.~Masuyama was an intern at MERL.}}}}
\address{$^1$Mitsubishi Electric Research Laboratories (MERL), Cambridge, MA, USA\\
$^2$Tokyo Metropolitan University, Tokyo, Japan}

\begin{document}
\ninept
\maketitle
\begin{abstract}

Head-related transfer functions (HRTFs) are important for immersive audio, and their spatial interpolation has been studied to upsample finite measurements.
Recently, neural fields (NFs) which map from sound source direction to HRTF have gained attention.
Existing NF-based methods focused on estimating the magnitude of the HRTF from a given sound source direction, and the magnitude is converted to a finite impulse response (FIR) filter.
We propose the neural infinite impulse response filter field (NIIRF) method that instead estimates the coefficients of cascaded IIR filters.
IIR filters mimic the modal nature of HRTFs, thus needing fewer coefficients to approximate them well compared to FIR filters.
We find that our method can match the performance of existing NF-based methods on multiple datasets, even outperforming them when measurements are sparse.
We also explore approaches to personalize the NF to a subject and experimentally find low-rank adaptation to be effective.

\end{abstract}
\begin{keywords}
Head-related transfer function, neural field, implicit neural representations, differentiable digital signal processing, parameter efficient fine-tuning
\end{keywords}
\section{Introduction}
\label{sec:intro}

Head-related transfer functions (HRTFs) capture the sound transmission from a sound source to  to both ears.
They are essential to many applications, including telepresence systems~\cite{Keyrouz2007} and virtual reality technologies~\cite{Xie2013,Serafin2018}.
As HRTFs depend on the shape of the pinnae, head, and upper torso, they vary from person to person, and individual HRTFs may lead to high-quality immersive audio experiences~\cite{Oberem2020,Andersen2021}.
When the individual HRTFs for all possible source directions are available, we can generate an immersive binaural signal by convolving the HRTFs with the source signals.
The ideal approach is to measure a sufficient number of HRTFs for each person, but it requires much time and effort because HRTFs are usually measured sequentially over several hundreds of directions~\cite{Watanabe2014}.
Hence, both spatial upsampling and personalization of HRTFs have been the focus of intense study~\cite{Gamper2013,Zotkin2003,Hu2008}.

To upsample the HRTF measurements, various techniques have been developed such as panning-based methods~\cite{Pulkki1997,Franck2017} and the spatial decomposition approach~\cite{Duraiswami2004,Ahrens2012,Arend2021}.
The former approach performs a weighted sum of the measured HRTFs for neighboring directions.
For instance, vector-base amplitude panning (VBAP)~\cite{Pulkki1997} has been widely used due to its low computational complexity, and it was adopted by the ISO/IEC MPEG-H 3D Audio standard~\cite{Herre2015}.
The latter approach expands the HRTFs to spatial basis functions, e.g., spherical harmonics~\cite{Ahrens2012,Arend2021}, and uses a global representation to estimate the HRTF for the target direction.
This approach can be interpreted as an optimization problem of the spatial coefficients~\cite{Duraiswami2004,Ahrens2012}, which becomes underdetermined when the number of measurements is limited.

Recently, the machine learning approach has gained increasing attention since this approach can flexibly exploit anthropometric features~\cite{Zhang2020,Lee2023} and HRTFs for multiple subjects~\cite{Ito2022,hogg2023,Gebru2021,Zhang2023}.
Despite recent progress, challenges remain regarding how to exploit a variable number of HRTF measurements and how to estimate HRTFs at arbitrary directions.
To tackle these challenges, several works have leveraged the neural field (NF), or implicit neural representation, where the HRTF is represented as a function of the sound source direction~\cite{Gebru2021,Zhang2023,Lee2023}.
NFs have been developed in computer vision to reconstruct 3D scenes from multiple 2D views~\cite{Mildenhall2022,Xie2022} and applied to spatial audio modeling~\cite{Luo2022,DiCarlo2023}.
In HRTF modeling by NFs, prior works have shown the potential to estimate the magnitude response of the HRTFs~\cite{Zhang2023,Lee2023}.
In these methods, the interpolated HRTFs are converted to time-domain finite impulse response (FIR) filters by the inverse discrete Fourier transform (DFT) with minimum phase~\cite{Kistler1992}.

\begin{figure}[t!]
\centering
\includegraphics[width=0.99\columnwidth]{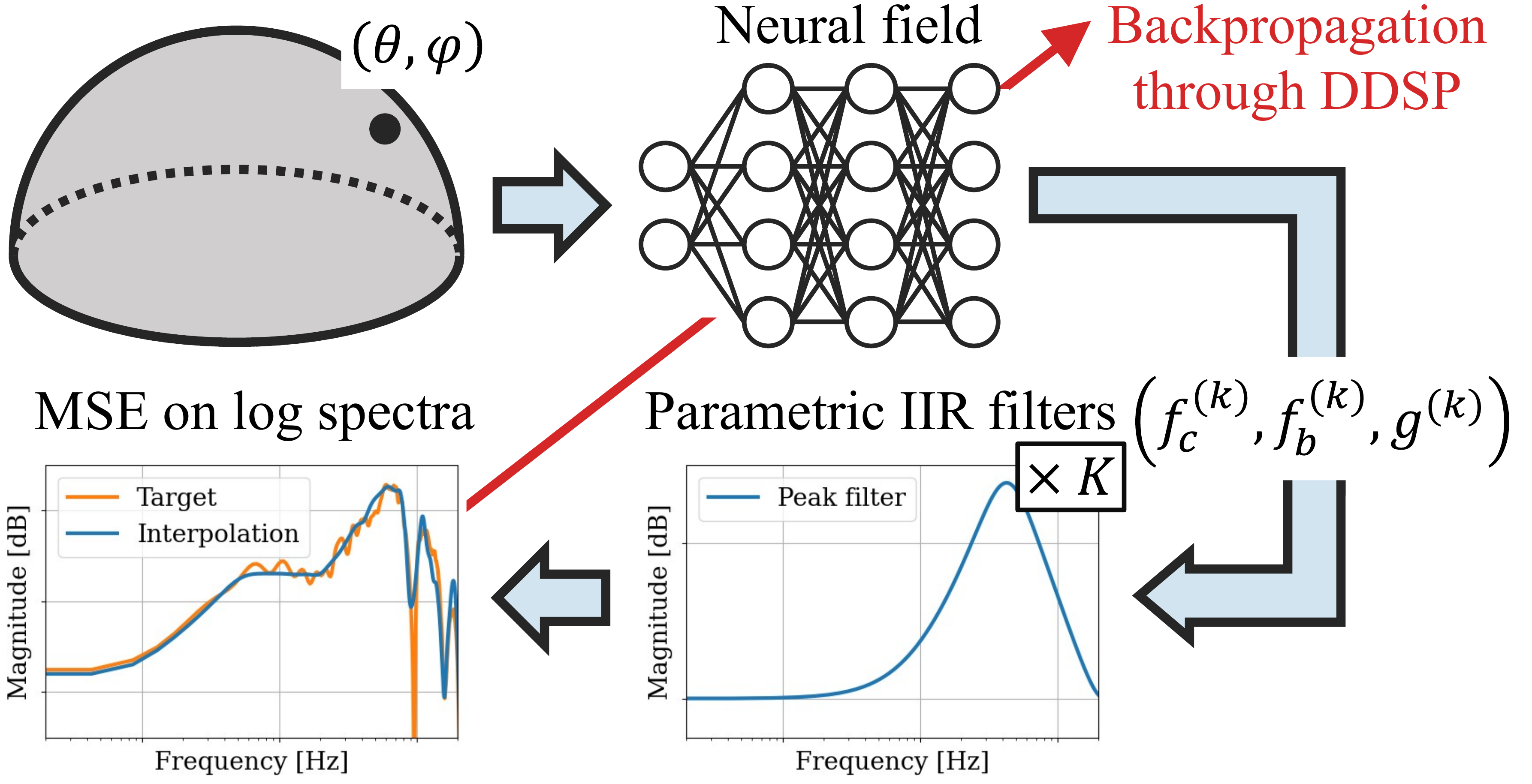}
\caption{Illustration of the proposed NF-based HRTF modeling method named NIIRF.
The cascade of IIR filters consist of $K$ peaking filters and two shelf filters for both left and right channels.
}
\vspace{-.3cm}
\label{fig:overview}
\end{figure}

Meanwhile, infinite impulse response (IIR) filters are advantageous for approximating HRTFs with fewer coefficients than FIR filters~\cite{Ramos2013,Nowak2022}.
This is preferable in terms of computational complexity and memory footprint.
While various algorithms are applicable to optimize the filter coefficients, back-propagation has recently been used to leverage powerful automatic differentiation libraries~\cite{Kuznetsov2020,Nercessian2020,Nercessian2021}.
This approach is in the family of differentiable digital signal processing (DDSP)~\cite{Engel2019} and has been applied to HRTF modeling~\cite{Bhattacharya2020}.
In addition, a neural network that estimates an IIR filter from a HRTF magnitude response was trained with DDSP~\cite{Colonel2022}.
These previous studies focused on approximating HRTFs only for certain directions and were not aimed at spatial upsampling.

In this paper, we propose to integrate NF-based spatial upsampling with differentiable IIR filters as a neural IIR filter field (NIIRF).
As depicted in Fig.~\ref{fig:overview}, our NF estimates the parameters of the cascaded IIR filters instead of the magnitude response or the time-domain FIR filter.
Building upon DDSP, we optimize the NF to minimize the error of the generated binaural filter responses by back-propagation.
Furthermore, we explore various conditioning approaches for personalizing the NF to new subjects.
Our experimental results confirm that the proposed method outperforms the classical panning-based baseline.
In addition, the proposed method improves the upsampling accuracy upon existing NF-based methods~\cite{Zhang2023}, when the number of measurements is limited.

\section{Proposed method: NIIRF}
\label{sec:prop}

\subsection{General formulation of HRTF modeling via NF}
\label{sec:general-form}

Defining $\theta$ as the azimuth in $[0, 2\pi)$ with $\theta=0$ in front of the subject, and $\phi$ as the elevation in $[-\pi/2, \pi/2]$ with $\phi=0$ on the equatorial plane, we aim to continuously interpolate HRTFs at any direction $(\theta, \phi)$. %
Each HRTF consists of a left-right pair of discrete-time filters, but our modeling does not include the interaural time difference (ITD) information, as our phase-insensitive loss allows for dealing with it independently similar to previous studies%
\footnote{The ITD can be estimated using several methods~\cite{Kistler1992,Ramos2013}.}%
~\cite{Ramos2013,Zhang2023}.

In practice, the neural field \texttt{NF} has been used to map directions to intermediary parameters $\mathbf{\Psi}_{\theta, \phi}$ that are then mapped onto an HRTF through a DSP algorithm.
This can be formulated as:
\begin{equation}
    \mathcal{H}_{\theta, \phi} = \texttt{DSP} \circ \texttt{NF} (\theta, \phi) = \texttt{DSP} (\mathbf{\Psi}_{\theta, \phi}),
    \label{eq:priorwork}
\end{equation}
where $\mathcal{H}_{\theta, \phi}$ is a filter representation of HRTF.
In prior works~\cite{Zhang2023,Lee2023}, $\mathbf{\Psi}_{\theta, \phi}$ is the magnitude of the DFT coefficients $|H_{\theta, \phi}[m]|$, where $m = 0, \ldots, M-1$ is the frequency bin index.
To train \texttt{NF} which estimates the magnitude response, we use the following mean squared error (MSE)~\cite{Zhang2023}:
\begin{equation}
    \mathcal{L} = \frac{1}{|\mathcal{D}|M} \sum_{(\theta, \phi) \in \mathcal{D}} \sum_{m=0}^{M-1} \left(20 \log_{10} \left| \frac{H_{\theta, \phi}[m]}{\widetilde{H}_{\theta, \phi}[m]} \right|\right)^2,
    \label{eq:mse}
\end{equation}
where $\widetilde{H}_{\theta, \phi}$ is the DFT of the measured HRTF, and $|\mathcal{D}|$ is the size of the dataset $\mathcal{D}$. 
In prior works following \eqref{eq:priorwork}, \texttt{DSP} is a phase estimation followed by an inverse DFT to estimate an FIR filter.
The phase is commonly recovered through minimum-phase reconstruction~\cite{Kistler1992}.

\subsection{Proposed integration of NF and DDSP}
\label{sec:overview}

Instead of estimating the DFT magnitude of the HRTF, we exploit an NF to estimate parameters of IIR filters as depicted in Fig.~\ref{fig:overview}.
This can be formulated as follows:
\begin{align}
    \mathcal{H}_{\theta, \phi} = \texttt{DDSP} \circ 
 \texttt{NF}_\mathrm{IIR}(\theta, \phi) = \texttt{DDSP}(\mathbf{\Psi}_{\theta, \phi}). \label{eq:iir-nf}
\end{align}
In the proposed method, $\mathcal{H}_{\theta, \phi}$ is obtained via a $\texttt{DDSP}$ module as a cascade of $K+2$ IIR filters for each ear, $\mathbf{\Psi}_{\theta, \phi}$ denoting a tuple of parameters for the IIR filters. %

\begin{figure}[t!]
\centering
\includegraphics[width=0.99\columnwidth]{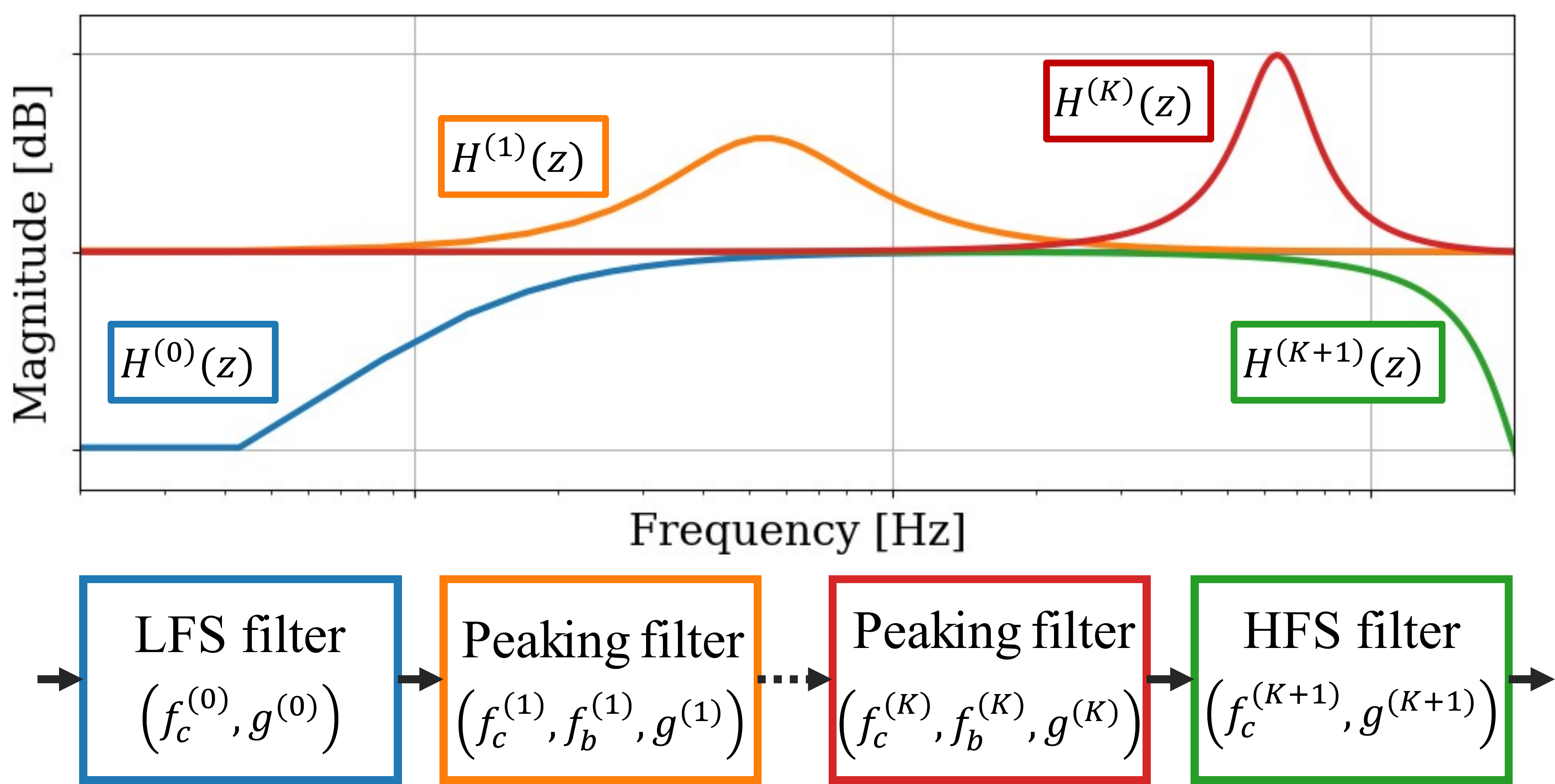}
\caption{Illustration of the cascade of parametric IIR filters.}
\label{fig:seqiir}
\end{figure}

More precisely, the \texttt{DDSP} module in \eqref{eq:iir-nf} converts the IIR parameters $\mathbf{\Psi}_{\theta, \phi}$ into a cascade of a first-order low-shelf (LFS) filter, $K$ second-order peaking filters, and a first-order high-shelf (HFS) filter as illustrated in Fig.~\ref{fig:seqiir}.
Using filter index $k = 0, \ldots, K+1$, the LFS and HFS filters are parameterized by their cut frequency $f_c^{(k)}$ in Hz and their gain $g^{(k)}$ in dB, and the peaking filters are parameterized by their center frequency $f_c^{(k)}$, their bandwidth $f_b^{(k)}$, and their gain $g^{(k)}$ following \cite{Bhattacharya2020,Zolzer2022}.
From these parameters, we can compute filter coefficients $b_n^{(k)}$ and $a_n^{(k)}$ as shown in the Appendix, where all operations are fully differentiable.
The transfer function of the $k$th filter is given by 
\begin{equation}
    H^{(k)}(z) =  \frac{b_0^{(k)} z^{0} + \cdots + b_N^{(k)} z^{-N}}{1 + \cdots + a_N^{(k)} z^{-N}},
\end{equation}
where $N=1$ for the shelf filters and $N=2$ for the peaking filters.
Instead of directly estimating the filter coefficients, we use parametric IIR filters to guarantee the stability of the estimated filters.

To train $\texttt{NF}_\mathrm{IIR}$ with the loss function in  \eqref{eq:mse}, we use the frequency-sampling method, i.e., uniformly sample the frequency axis into $M$ steps to get DFT coefficients as follows:
\begin{equation}
    H_{\theta, \phi}[m] = \prod_{k=0}^{K+1} H_{\theta, \phi}^{(k)}(z_m),
    \label{eq:frequency-sampling}
\end{equation}
where $z_m = \exp(2\pi j m / M)$, and $j$ is the imaginary unit%
\footnote{
DFT coefficients in \eqref{eq:frequency-sampling} technically correspond to a time-aliased version of the filter response~\cite{rabiner1970}.
That error can however be well mitigated by using a large enough $M$.
}.

Once $\texttt{NF}_\mathrm{IIR}$ is trained, we can compute the IIR filters for arbitrary directions and apply the filters in the time domain.
The proposed method has the potential to reduce the memory footprint because IIR filters can better model acoustic systems with fewer coefficients than FIR filters, such as those obtained through the existing NF-based method~\cite{Zhang2023}.
We also expect the parameters of the IIR filters to be easier to interpolate than the magnitude response.

\subsection{NF for estimating IIR filter parameters}
\label{sec:nf-for-iir}

For $\texttt{NF}_\mathrm{IIR}$ in \eqref{eq:iir-nf}, we use a multi-layer perceptron (MLP) with the GeLU activation functions on the random Fourier features (RFF)~\cite{Tancik2020}.
In detail, the sound source direction $(\theta, \phi)$ is converted to the RFF as follows:
\begin{equation}
    \!\! \mathbf{e}
    = [\cos([\mathbf{p}_0^\mathsf{T} \mathbf{d}), \sin(\mathbf{p}_0^\mathsf{T} \mathbf{d}), \ldots, \cos(\mathbf{p}_{C-1}^\mathsf{T} \mathbf{d}), \sin(\mathbf{p}_{C-1}^\mathsf{T} \mathbf{d})]^\mathsf{T}, \!\!
    \label{eq:rff}
\end{equation}
where $\mathbf{d} = [\theta - \pi, \phi]^\mathsf{T}$, $\mathbf{p}_c \in \mathbb{R}^2$ is sampled from an isotropic Gaussian distribution, $c=0, \ldots, C-1$ is the channel index, and $(\cdot)^\mathsf{T}$ denotes the transpose.
In our preliminary experiments, the use of RFF stabilized the learning curve compared with SIREN~\cite{Sitzmann2020} and slightly improved the performance from the vanilla positional encoding~\cite{Mildenhall2022}.

At the last layer of the MLP, we employ the sigmoid function to control the frequency range of $f_c^{(k)}$ as follows:
\begin{equation}
    f_c^{(k)}(x_c^{(k)}) = (f_\text{max}^{(k)} - f_\text{min}^{(k)}) \sigmoid(x_c^{(k)}) + f_\text{min}^{(k)},
\end{equation}
where $x_c^{(k)}$ is an element of the output of the last affine layer.
We design ranges $[f_\text{min}^{(k)}, f_\text{max}^{(k)}]$ by first finding the frequencies of all peaks and notches in the magnitude response of all HRTFs in a training dataset, sorting these frequencies, and using quantiles to split the interval in equally distributed ranges while ensuring neighboring ranges have 50\% overlap.
By assigning different frequency ranges to each of the $K$ peak filters, we avoid having a filter cascade with several peaking filters clustered around the same spectral peak.
Meanwhile, the range of $f_b^{(k)}$ is fixed as
\begin{equation}
    f_b^{(k)}(x_b^{(k)}) = (\tilde{f}_\text{max} - \tilde{f}_\text{min}) \sigmoid(x_b^{(k)}) + \tilde{f}_\text{min},
\end{equation}
where $x_b^{(k)}$ is another element of the last affine layer's output.
We do not use any activation functions to obtain $g^{(k)}$ as it is unconstrained.

\subsection{Efficient adaption of NF to new subjects}
\label{sec:multi-people}

We have so far described an NF for a single subject.
As HRTFs are similar across different subjects, we expect that an NF with a small number of subject-specific parameters can efficiently represent the HRTFs of multiple subjects, and define
\begin{equation}
    \mathbf{\Psi}_{\theta, \phi} = \texttt{NF}_\mathrm{IIR}(\theta, \phi \mid i), \label{eq:iir-bf-multi}
\end{equation}
where $i$ is introduced as a subject index to change subject-specific parameters.
A popular approach to conditioning NFs is \mbox{conditioning} by concatenation (CbC)~\cite{Xie2022}.
CbC introduces a subject embedding $\mathbf{z}_i$ and concatenates it to the model input; it has been used in HRTF spatial upsampling~\cite{Zhang2023}.
Another approach is feature-wise linear modulation (FiLM)~\cite{Perez2018}, which uses another neural network to compute modulation parameters $(\boldsymbol{\mu}_{l,i}, \boldsymbol{\sigma}_{l,i})$ for each layer $l$ from $\mathbf{z}_i$.
The output of layer $l$ is then modified from $\gelu (\mathbf{A}_l \mathbf{x}_l + \mathbf{b}_l)$ to
\begin{equation}
    \texttt{FiLM}(\mathbf{x}_l \mid i) = \boldsymbol{\sigma}_{l, i} \odot \gelu (\mathbf{A}_l \mathbf{x}_l + \mathbf{b}_l) + \boldsymbol{\mu}_{l, i},
    \label{eq:film}
\end{equation}
where $\odot$ denotes the Hadamard product, $\mathbf{x}_l$ is the input of the layer $l$, and $\mathbf{A}_l$ and $\mathbf{b}_l$ are parameters of the MLP shared across subjects.

The subject embedding $\mathbf{z}_i$ can be optimized by back-propagation similar to other parameters, a method referred to as auto-decoding \cite{Xie2022}.
To personalize the NF to a new subject, we optimize only $\mathbf{z}_i$ to minimize the MSE in \eqref{eq:mse} while freezing the shared parameters.
Auto-decoding can easily handle an arbitrary number of measurements of the target subject because it does not need additional neural networks to extract $\mathbf{z}_i$ from the measurements.

We also explore two parameter-efficient fine-tuning methods: bias-terms fine-tuning (BitFit)~\cite{Zaken2021} and low-rank adaptation (LoRA)~\cite{Hu2021}.
BitFit adapts a neural network to a new domain by fine-tuning the bias term $\mathbf{b}_l$ instead of all parameters.
Here, we introduce a different bias term $\mathbf{b}_{l,i}$ for each subject: $\texttt{BitFit}(\mathbf{x}_l \mid i)  = \gelu (\mathbf{A}_l \mathbf{x}_l +\mathbf{b}_{l,i}).$
LoRA introduces an additional rank-$1$ matrix for each layer as a product of two vectors%
\footnote{In general, LoRA can also leverage low-rank matrices whose rank is higher than $1$. In this paper, we focus on the case with rank-$1$ for simplicity.}%
:
\begin{equation}
    \texttt{LoRA}(\mathbf{x}_l \mid i)  = \gelu (\mathbf{A}_l \mathbf{x}_l + \mathbf{u}_{l,i} \mathbf{v}_{l,i}^{\mathsf{T}} \mathbf{x}_l +  \mathbf{b}_l),
\end{equation}
where $\mathbf{u}_{l,i}$ and $\mathbf{v}_{l,i}$ are the subject-specific vectors for building the rank-$1$ matrix.
After optimizing the rank-$1$ matrix for each target subject, we can add it to the original weight $\mathbf{A}_l$, which reduces the computational cost to obtain the HRTFs for the same subject.

\section{Experiments}
\label{sec:exp}

\subsection{Comparison with existing single-subject methods}
\label{sec:exp-single}

\begin{figure}[t!]
\centering
\includegraphics[width=0.99\columnwidth]{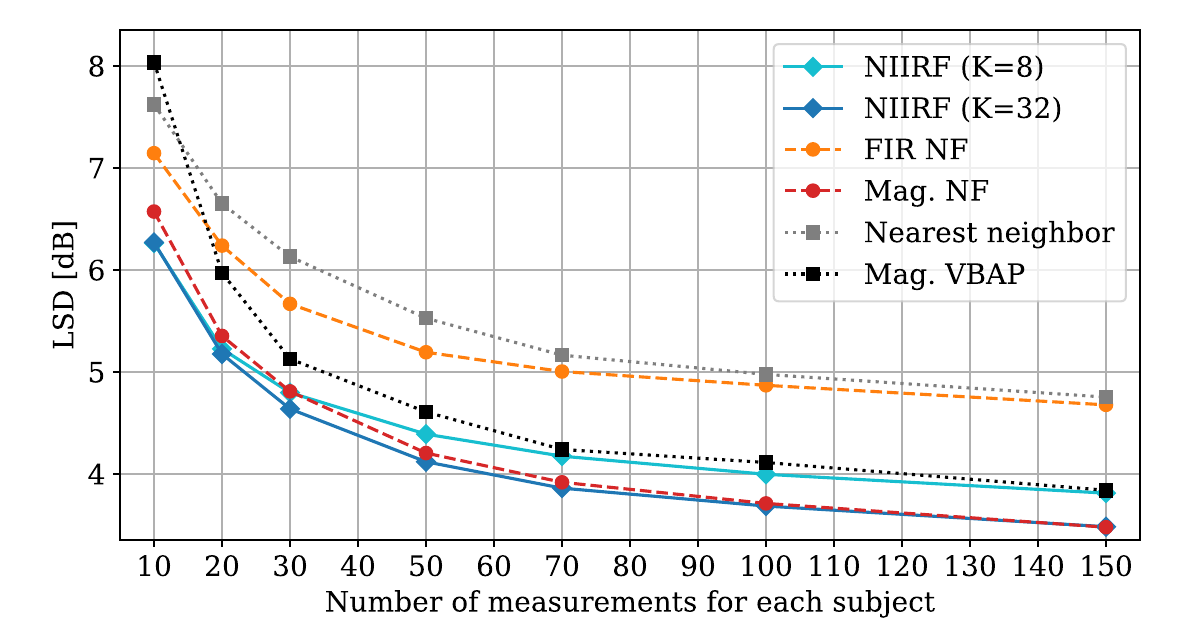}
\vspace{-.3cm}
\caption{LSD with different number of measurements for training.}
\label{fig:lsd-single}
\vspace{-.3cm}
\end{figure}

We first validate the interpolation accuracy of the proposed method under a single-subject setting.
We used the CIPIC dataset~\cite{Algazi2001},~which includes $45$ subjects and $1250$ measurements per subject.
The measurements were randomly split into three sets: $1000$ for evaluation, $100$ for validation, and $150$ for training.
The sampling frequency was $44.1$ kHz, and we used the DFT with $512$ points to compute $\widetilde{H}_{\theta, \phi}[m]$ in \eqref{eq:mse}.
The log-spectral distortion (LSD), defined as the square root of $\mathcal{L}$ in \eqref{eq:mse} measured from $20$ Hz to $20$ kHz, was used to perform evaluation on the audible frequency range (lower is better).

Our NF (``NIIRF'') consists of four hidden layers with $512$ units and an output layer whose size depends on the number of peaking filters $K$.
The NF was trained using the RAdam optimizer~\cite{liu2020} with a learning rate of $5 \times 10^{-4}$.
We used the model with the best loss $\mathcal{L}$ on the validation set for evaluation.
We compared the proposed method with two baselines: the nearest neighbor algorithm (``Nearest neighbor'') and a variant of VBAP~\cite{Franck2017} (``Mag.\ VBAP'').
In the latter baseline, we performed a weighted sum of the magnitude responses in the training dataset, where the weight is calculated following~\cite{Franck2017}.
NF-based methods estimating the magnitude response~\cite{Zhang2023} (``Mag.\ NF'') and the time-domain FIR filter (``FIR NF'') were also evaluated.
They differ from the proposed method only by the output layer.

Figure~\ref{fig:lsd-single} illustrates the variations in the LSD averaged over $45$ subjects depending on the number of HRTF measurements available for training.
The proposed method with $8$ filters achieved comparable performance with ``Mag.\ VBAP,'' even outperforming it when measurements are sparse.
The NF-based method for the FIR filter performed poorly compared with that for the magnitude response.
This is likely because variation in time domain impulse responses across different directions is much harder to approximate with a fixed model size compared to variations in their spectra.
As a result of changing the estimation target to the parameters of the IIR filters, the proposed method with $32$ peaking filters slightly improved the performance when the number of HRTF measurements was limited.

\begin{figure}[t!]
\centering
\includegraphics[width=0.99\columnwidth]{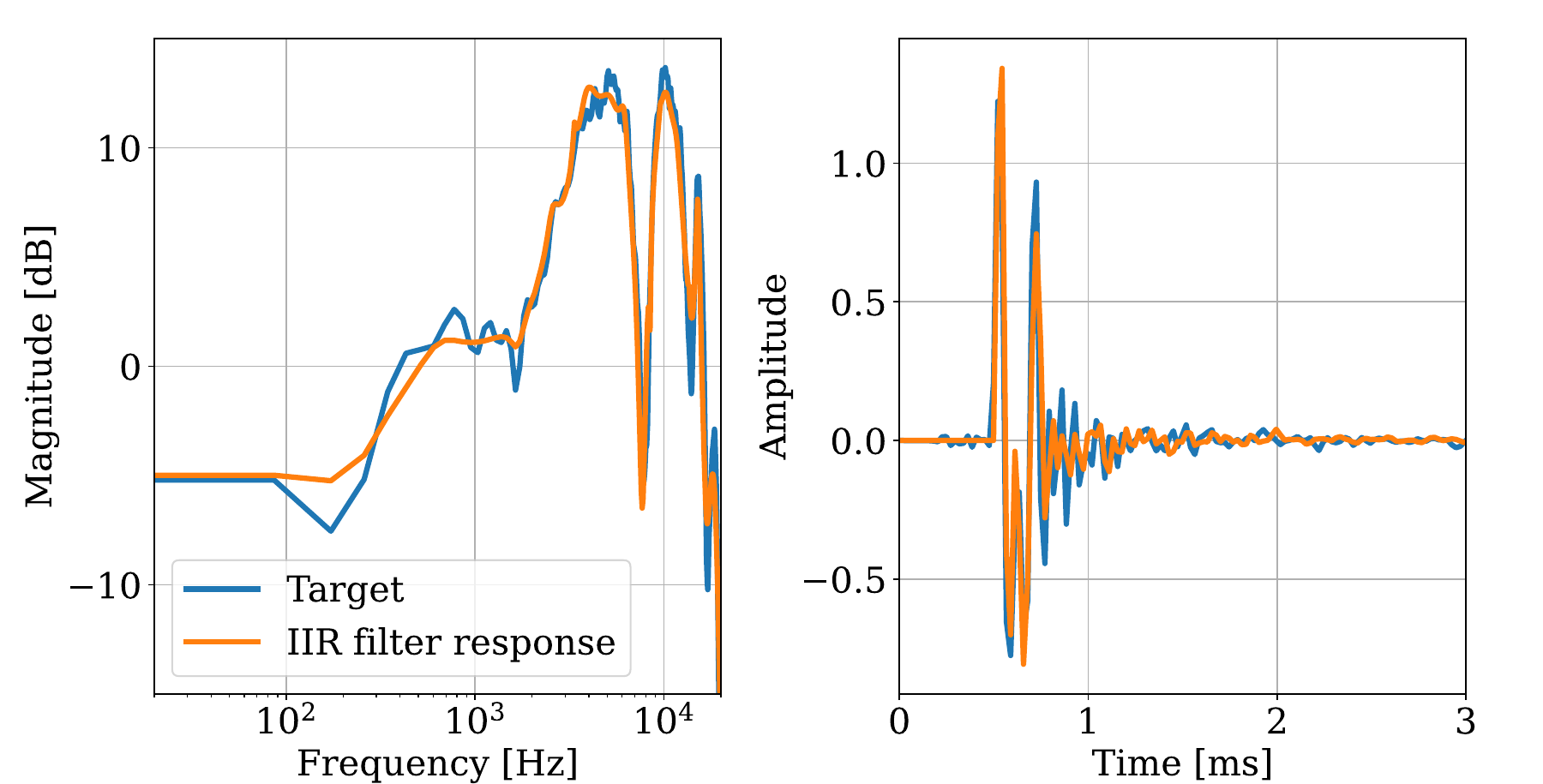}
\vspace{-.3cm}
\caption{Estimated filter response in the frequency and time domains.}
\label{fig:examples}
\end{figure}

Figure~\ref{fig:examples} shows examples of the responses of the left-channel IIR filter in the frequency and time domains, where we used the model trained for subject $3$ with $150$ measurements%
\footnote{
In Fig.~\ref{fig:examples}, we shift the estimated filter response so that its correlation with the target impulse response is maximized, which corresponds to compensating for the transmission delay from a sound source to the ear~\cite{Ramos2013}.
Several methods have been developed to interpolate the delay~\cite{Brinkmann2015}.
}.
The time-domain filter response captured the shape of the head-related impulse response even though the NF was trained on the phase insensitive loss in \eqref{eq:mse}.
This is a favorable feature of the cascaded IIR filters~\cite{Ramos2013,Nowak2022}.

\subsection{Adaptation of multi-subject NF to new subjects}

We use the HUTUBS dataset~\cite{Brinkmann2019} for evaluating adaptation of an NF pre-trained on multiple subjects to new subjects. The dataset contains $94$ subjects excluding two repeated subjects~\cite{Ito2022}, and $440$ measurements per subject. The first $87$ subjects were used for pre-training, and the remaining $7$ subjects were used for evaluating the adaptation capability.  $100$ randomly selected directions are held out from all $94$ subjects, and used to form an unseen direction test set (``Test2’’). An additional $100$ directions are held out from the $7$ adaptation subjects to form a seen direction test set (``Test 1’’), and also held out from $10$ of the training subjects for use as a validation set.  All remaining directions are used for either pre-training or adaptation.
The NFs were pre-trained on the HRTFs of multiple subjects using the RAdam optimizer, and then the subject-specific parameters were optimized for each target subject with the AdamW optimizer.

\begin{table}[t]
    \centering
    \sisetup{
    detect-weight, %
    mode=text, %
    tight-spacing=true,
    round-mode=places,
    round-precision=1,
    table-format=1.1,
    table-number-alignment=center
    }
    \vspace{-.1cm}
    \caption{LSD [dB] for different numbers of measurements for adaptation. The number of the subject-specific parameters adapted to the target subject is indicated by $Q$.}
    \vspace{-.2cm}
    \resizebox{\linewidth}{!}{%
    \begin{tabular}[t]{cccccccc}
    \toprule
   &&&\multicolumn{5}{c}{Number of measurements} \\
    \cmidrule(lr){4-8}
     Method & Adaptation &$Q$ & 10 & 20 & 30 & 50 & 100 \\
    \midrule
    \multicolumn{8}{c}{\emph{Directions seen in pre-training (Test1)}} \\
    \midrule
    \multirow{4}{*}{Mag.\ NF}
    & CoC~\cite{Zhang2023} & 32
    & 4.8 & 4.7 & 4.6 & 4.6 & 4.5 \\
    & FiLM & 32
    & \bf{4.3} & 4.2 & 4.2 & 4.1 & 4.1 \\
    & BitFit & 2562
    & \bf{4.3} & \bf{4.0} & 3.9 & 3.7 & \bf{3.5} \\
    & LoRA & 5122
    & \bf{4.3} & \bf{4.0} & \bf{3.8} & \bf{3.6} & \bf{3.5} \\
    \midrule
    \multirow{4}{*}{\shortstack[c]{NIIRF ($K=32$) \\ Proposed}}
    & CoC & 32
    & 4.8 & 4.7 & 4.7 & 4.6 & 4.6 \\
    & FiLM & 32
    & 4.3 & 4.2 & 4.2 & 4.2 & 4.2 \\
    & BitFit & 2248
    & \bf{4.3} & \bf{4.0} & 3.9 & 3.7 & \bf{3.5} \\
    & LoRA & 4808
    & \bf{4.3} & \bf{4.0} & 3.9 & 3.7 & \bf{3.5} \\
    \midrule
    \multicolumn{8}{c}{\emph{Directions unseen completely (Test2)}} \\
    \midrule
    \multirow{4}{*}{Mag.\ NF} 
    & CoC~\cite{Zhang2023} & 32
    & 4.9 & 4.8 & 4.8 & 4.8 & 4.7  \\
    & FiLM & 32
    & \bf{4.5} & \bf{4.4} & 4.4 & 4.4 & 4.3 \\
    & BitFit & 2562
    & 5.0 & 4.8 & 4.6 & 4.4 & 4.4 \\
    & LoRA & 5122
    & 5.2 & 5.0 & 4.9 & 4.8 & 4.6 \\
    \midrule
    \multirow{4}{*}{\shortstack[c]{NIIRF (K=32) \\ Proposed}} 
    & CoC & 32
    & 5.0 & 4.9 & 4.9 & 4.8 & 4.7 \\
    & FiLM & 32
    & \bf{4.5} & 4.5 & 4.4 & 4.4 & 4.4 \\
    & BitFit & 2248
    & 4.8 & 4.5 & 4.4 & 4.2 & 4.1 \\
    & LoRA & 4808
    & 4.7 & \bf{4.4} & \bf{4.2} & \bf{4.1} & \bf{4.0} \\
    \bottomrule
    \end{tabular}}
    \vspace{-.3cm}
\label{table:lsds}
\end{table}

Table~\ref{table:lsds} presents the LSD with different adaptation methods.
As a point of comparison, we trained a single-subject model as used in Section~\ref{sec:exp-single} for each of the $7$ target subjects.
For the directions included in the multi-subject pre-training (Test1), the average performance over all subjects was $7.5$ dB when training on $10$ measurements, and $5.1$ dB when training on $100$ measurements. %
We see from Table~\ref{table:lsds} that all the adaptation methods are effective for HRTF personalization and dramatically improve upon these results.
NF-based methods estimating the magnitude response and the parametric IIR filters resulted in a similar performance on Test1, with BitFit and LoRA performing best.
However, the proposed NIIRF with LoRA outperformed the existing NF-based method in most cases on Test2.
We hypothesize that generalization is better for NIIRF because values specifying parametric filters such as filter center frequencies and gains are more predictable across directions than spectra.

\vspace{-.1cm}
\section{Conclusion}
\label{sec:conclusion}
\vspace{-.1cm}

We proposed to integrate an NF approach and a cascade of differentiable IIR filters for HRTF modeling.
The NF is trained to minimize the error of the magnitude response based on DDSP.
We demonstrate that the proposed method improves the upsampling accuracy when the number of HRTF measurements is limited. 
We also investigate various adaptation methods to personalize the NF to a new subject.
Our source code is available online%
\footnote{
\url{https://github.com/merlresearch/neural-IIR-field}
}.

\vspace{-.1cm}
\section{Appendix: IIR Coefficient Computation}
\vspace{-.1cm}
\label{sec:appendix}

Following \cite{Bhattacharya2020,Zolzer2022}, the transfer functions of the first-order LFS and HFS filters are defined as in Table~\ref{table:shelf}, where $\eta^{(k)} = (\rho^{(k)}-1) / 2$, $\rho^{(k)} = 10^{g^{(k)}/20}$, and $f_s$ is the sampling rate.
The transfer function of the second-order peaking filter is summarized in Table~\ref{table:peaking}.

\begin{table}[h]
    \centering
    \sisetup{
    detect-weight, %
    mode=text, %
    tight-spacing=true,
    round-mode=places,
    round-precision=1,
    table-format=1.1,
    table-number-alignment=center
    }
    \vspace{-.5cm}
    \caption{Coefficients of the first-order shelf filters.}
    \vspace{-.2cm}
    \resizebox{.9\linewidth}{!}
    {
    \begin{tabular}[t]{ccc}
    \toprule
    & LFS & HFS \\
    \midrule
    $b_{0}^{(k)}$
    & $1 + \eta^{(k)} (1 + \alpha^{(k)})$
    & $1 + \eta^{(k)} (1 - \alpha^{(k)})$ \\
    $b_{1}^{(k)}$
    & $\alpha^{(k)} + \eta^{(k)} (\alpha^{(k)} + 1)$
    & $\alpha^{(k)} + \eta^{(k)} (\alpha^{(k)} - 1)$ \\
    $a_{1}^{(k)}$ 
    & $\alpha^{(k)}$
    & $\alpha^{(k)}$ \\
    \midrule
    \begin{tabular}{c}$\alpha^{(k)}$ \\ if $g^{(k)} \geq 0$ \end{tabular}
    & 
    $\frac{\tan(\pi f_c^{(k)} / f_s) - 1}{\tan(\pi f_c^{(k)} / f_s) + 1}$
    & $\frac{\tan(\pi f_c^{(k)} / f_s) - 1}{\tan(\pi f_c^{(k)} / f_s) + 1}$ \\
    \midrule
    \begin{tabular}{c}$\alpha^{(k)}$ \\ if $g^{(k)} < 0$ \end{tabular}
    & 
    $\frac{\tan(\pi f_c^{(k)} / f_s) - \rho^{(k)}}{\tan(\pi f_c^{(k)} / f_s) + \rho^{(k)}}$
    & $\frac{ \rho^{(k)}\tan(\pi f_c^{(k)} / f_s) - 1}{ \rho^{(k)}\tan(\pi f_c^{(k)} / f_s) + 1}$ \\
    \bottomrule
    \end{tabular}}
    \vspace{-.4cm}
\label{table:shelf}
\end{table}

\begin{table}[h]
    \centering
    \sisetup{
    detect-weight, %
    mode=text, %
    tight-spacing=true,
    round-mode=places,
    round-precision=1,
    table-format=1.1,
    table-number-alignment=center
    }
    \vspace{-.4cm}
    \caption{Coefficients of the second-order peaking filters.}
    \vspace{-.2cm}
    \resizebox{.7\linewidth}{!}
    {
    \begin{tabular}[t]{cc}
    \toprule
    & Peaking filter \\
    \midrule
     $b_0^{(k)}$
    & $1+ \eta^{(k)} (1 + \beta^{(k)})$ \\
    $b_1^{(k)}$
    & $\gamma^{(k)} (1 - \beta^{(k)})$ \\
    $b_2^{(k)}$
    & $-\beta^{(k)} - \eta^{(k)} (1+ \beta^{(k)})$ \\
    $a_1^{(k)}$
    & $\gamma^{(k)} (1 - \beta^{(k)})$ \\
    $a_2^{(k)}$
    & $-\beta^{(k)}$ \\
    \midrule
    \begin{tabular}{c}$\beta^{(k)}$ \\ if $g^{(k)} \geq 0$ \end{tabular}
    & $\frac{\tan(\pi f_b^{(k)} / f_s) - 1}{\tan(\pi f_b^{(k)} / f_s) + 1}$ \\
    \midrule
    \begin{tabular}{c}$\beta^{(k)}$ \\ if $g^{(k)} < 0$ \end{tabular}
    & $\frac{\tan(\pi f_b^{(k)} / f_s) - \rho^{(k)}}{\tan(\pi f_b^{(k)} / f_s) + \rho^{(k)}}$ \\
    \midrule
    $\gamma^{(k)}$
    & $- \cos\left( 2 \pi \frac{f_c^{(k)}}{f_s} \right)$ \\
    \bottomrule
    \end{tabular}}
    \vspace{-.4cm}
\label{table:peaking}
\end{table}

\clearpage
\balance
\footnotesize
\bibliographystyle{IEEEbib}
\bibliography{refs}

\end{document}